\documentclass[reprint,prb,twocolumn,showpacs,superscriptaddress]{revtex4-1}

\usepackage{graphicx}
\DeclareGraphicsExtensions{.png,.jpg,.eps}

\usepackage{amsmath}

\begin{document}

\title{Non-collinear vs collinear description of the Ir-based one-t$_{2g}$-hole perovskite-related compounds: SrIrO$_3$ and Sr$_2$IrO$_4$}

\author{J. L. Lado}
\affiliation{International Iberian Nanotechnology Laboratory, Braga, Portugal}  
  \email{jose.lado@inl.int}
	
\author{V. Pardo}
\affiliation{Departamento de F\'{i}sica Aplicada,
  Universidade de Santiago de Compostela, E-15782 Campus Sur s/n,
  Santiago de Compostela, Spain}
\affiliation{Instituto de Investigaci\'{o}ns Tecnol\'{o}xicas,
  Universidade de Santiago de Compostela, E-15782 Campus Sur s/n,
  Santiago de Compostela, Spain}  
  \email{victor.pardo@usc.es}


\begin{abstract} 

We present an analysis of the electronic structure of perovskite-related iridates,
5d electron compounds where a subtle interplay between spin-orbit
coupling, tetragonal distortions and electron correlations determines the electronic structure properties. We suggest
via electronic
structure calculations
that a non-collinear calculation is required to obtain solutions close
to the usually quoted j$_{eff}$= 1/2 state to describe the t$_{2g}$ hole in the Ir$^{4+}$:d$^5$ cation, while a collinear calculation yields
a different solution, the hole is in a simpler $xz/yz$ complex
combination with a smaller $L_z/S_z$ ratio.
We describe what the implications of
this are in terms of the electronic
structure;  surprisingly, both solutions barely differ
in terms of their band structure, and are similar to the
one obtained by a tight binding model involving t$_{2g}$ orbitals
with mean field interactions.
We also analyze how the electronic structure and magnetism evolve with 
strain, spin-orbit coupling strength and on-site
Coulomb repulsion; we suggest the way the band structure
gets modified and draw some comparisons with
available experimental observations.
 
\end{abstract}
\maketitle


\section{Introduction}

A significant amount of work has been developed recently on the interesting physical properties
of perovskite-based iridates. These are Ir$^{4+}$:d$^5$ based compounds with structure related to that of the perovskite (metal cations surrounded by oxygen octahedra form the local environment) where dimensionality can be tuned by moving along the Ruddelsden-Popper series A$_{n+1}$Sr$_{n}$O$_{3n+1}$.\cite{rud_pop_1,rud_pop_2} Octahedral crystal field together with large spin-orbit coupling produce one hole in the t$_{2g}$ band, forming a j$_{eff}$= 1/2 state that could somehow resemble the situation in cuprates: a single hole in a square lattice, with antiferromagnetic (AF) correlations dominating, but the t$_{2g}$ manifold is involved in this case instead of the $e_g$ that occurs in cuprates. Based on these similarities, the appearance of superconductivity has been speculated,\cite{iridates_sc} and recently experimental evidences have started to emerge suggesting this could indeed be the case,\cite{sr2iro4_sc_1,sr2iro4_sc_2} or at least that some Fermi surface features are very similar to underdoped cuprates.\cite{sr2iro4_delatorre} These systems are also interesting because the role of spin-orbit coupling in the electronic structure is not completely clear, in particular it has been under debate what causes the insulating behavior in Sr$_2$IrO$_4$, whether the system is a Mott or a Slater insulator.\cite{sr2iro4_novel_soc_Mott,sr2iro4_slater,sr2iro4_slater_prb14,sr2iro4_mott_slater_mixture,sr2iro4_mott_slater_mixture_prb12,sr2iro4_novel_soc_Mott,sr2iro4_slater_scirep} One of the keys to answer this question is to describe the electronic state the one t$_{2g}$ hole is in. This has been described as a pure j$_{eff}$= 1/2 state\cite{sr2iro4_jeff_1_2,sr2iro4_slater} or as a mixture due to contributions from both j$_{eff}$= 3/2 and e$_g$ states.\cite{sr2iro4_mixed,sriro3_mixed,sr2iro4_sr2rho4_lda_dmft,sr2iro4_pressure} To elucidate this issue, several iridates have been measured via different spectroscopic techniques, revealing that the L$_z$/S$_z$ ratio can be different\cite{sr2iro4_lz_sz,bairo3_prl10} from the expected value of 4 that would occur for the pure j$_{eff}$= 1/2 state.\cite{sr2iro4_jeff_1_2} At present it is not clear what the actual size of the different parameters involved in determining the electronic structure should be (Hubbard U, spin-orbit coupling strength, band width, etc.), thus a careful study of the evolution of the electronic structure with these could shed light on the electronic structure properties of this family of compounds. 

Thin films of these iridates have been grown under different conditions. With them, various experiments on the effects of strain in modifying their electronic structure have been carried out. Both SrIrO$_3$ and Sr$_2$IrO$_4$ have been studied under various types of compressive and tensile strain. The results indicate that SrIrO$_3$, being a paramagnetic metal in the bulk, can become insulating via both disorder-induced Anderson-type localization or by the application of strain.\cite{sriro3_mit} In Sr$_2$IrO$_4$ a similar situation occurs, the charge gap gets broadened by the application of compressive strain when compared to the bulk case,\cite{sr2iro4_gap_vs_strain_2} but the X-ray absorption spectra (XAS) obtained in films grown on different substrates shows the opposite behavior,\cite{sr2iro4_gap_vs_strain} that the gap increases under tensile strain. There is a controversy here that we will try to address from electronic structure calculations.

\begin{figure}[h!]
\begin{center}
\includegraphics[width=0.9\columnwidth,draft=false]{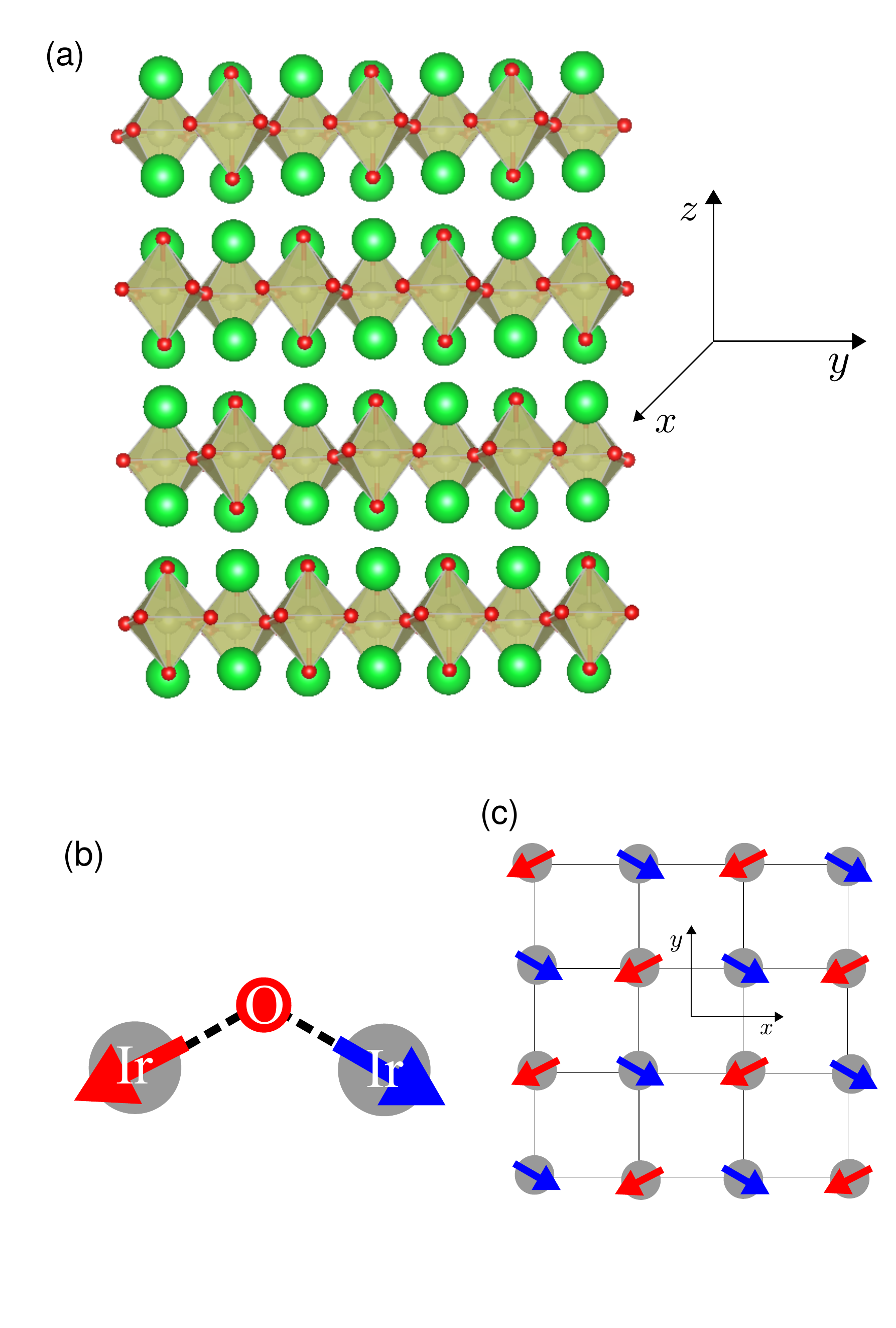}
\caption{(Color online.) (a) Layered structure of Sr$_2$IrO$_4$, where the oxygen octahedral environment around the Ir atoms is highlighted. (b) Sketch of an Ir-O-Ir bond, with
the (exaggerated) canted magnetic order
between Ir atoms, which arises due to the
interplay of spin-orbit coupling and the non-linear hopping path through oxygens.
(c) Schematic top view of the in-plane antiferromagnetic ordering of the Ir atoms.
The net magnetic moments lie on the plane of the layered square lattice, slightly canted towards the neighboring oxygen, leading to a weak FM component.}
\label{sr2iro4}
\end{center}
\end{figure}

Sr$_2$IrO$_4$ is a magnetic insulator that shows a positive $\Theta_{CW}$ with a small ordered moment compared to the effective paramagnetic moment that has been measured.\cite{sr2iro4_mu_eff} Some papers indicate the ordered moment can be produced via slight canting of the otherwise AF moments.\cite{sr2iro4_canted_af_science_09,sr2iro4_canted_af_prl12_a,sr2iro4_canted_af_prl12_b, sr2iro4_canted_af_prb13} The canting angle (see Fig. \ref{sr2iro4}b) occurs due to the deviation of the Ir-O-Ir angle from 180$^{\circ}$ via Dzyaloshinskii-Moriya interaction.\cite{dzyalo,moriya} It is to be understood if this canted AF solution is consistent with the positive $\Theta_{CW}$ observed.\cite{sr2iro4_theta_positive} We will show ab initio from calculations including a non-collinear description of magnetism that such a canting occurs naturally due to the structural distortion.

We provide here an electronic structure analysis of both SrIrO$_3$ and Sr$_2$IrO$_4$. We have analyzed the evolution of their electronic structures with the application of strain, pressure, varying the on-site Coulomb repulsion, analyzing the role of spin-orbit coupling and comparing a non-collinear and a collinear description for magnetism. 
Surprisingly, we find that non-collinearity is crucial in
determining the symmetry and character
of the t$_{2g}$ hole, but the band structure of the system
is barely affected by this orbital character.
In the particular case of Sr$_2$IrO$_4$, we show how a tight binding model
captures many important features of the behavior of the system,
yielding also similar band structures even without including the
canting of local moments or octahedral rotations.
We have tried to elucidate in what limit of spin-orbit coupling
vs band width and vs U the system is and how one can
understand the properties of the ground state of these compounds. 


\section{Computational procedures}


Ab initio electronic structure calculations based on the density functional theory (DFT)\cite{hk,ks} have been performed
using two all-electron full potential codes ({\sc wien2k}\cite{wien} and {\sc Elk}\cite{elk}) on various Ir compounds, whose structure will be discussed below.
The exchange-correlation term is parametrized depending
on the case. We have used the generalized gradient
approximation
(GGA) in the Perdew-Burke-Ernzerhof\cite{gga} scheme for structural optimizations (atomic positions and volume optimizations). For treating on-site Coulomb repulsion in the 5d manifold, we have employed the
local density approximation+U
(LDA+U) method in various rotational invariant flavors: in the so-called ``fully localized limit",\cite{sic} and
for the {\sc elk} calculations also
with the Yukawa scheme.\cite{yukawa}  A
non-collinear scheme for treating the magnetic moments
was also used within the {\sc Elk} code, and compared with the collinear formalism utilizing the same code.

Regarding the {\sc wien2k} calculations,
calculations were performed with a converged k-mesh and a value of  R$_{mt}$K$_{max}$= 7.0.
Spin-orbit coupling (SOC) was introduced in a second variational manner using the scalar
relativistic approximation.\cite{singh}
The $R_{mt}$ values used were in a.u.: 2.23 for Sr, 1.96 for Ir and 1.60 for O when studying Sr$_2$IrO$_4$, and 2.50 for Sr, 2.00 for Ir and 1.63 for O when analyzing SrIrO$_3$.

We have carried out calculations in SrIrO$_3$ and Sr$_2$IrO$_4$, using the structures from Refs. \onlinecite{sriro3_struct} and \onlinecite{sr2iro4_struct}, respectively. We have simulated the effects of both tensile and compressive strains by fixing the $a$ lattice parameter to that of several well-known systems typically used as substrates for thin film deposition (KTaO$_3$: $a$= 3.989 \AA, LSAT: $a$=3.868 \AA, SrTiO$_3$: $a$= 3.905 \AA, LaSrAlO$_4$: $a$= 3.755 \AA, MgO: $a$= 4.212 \AA) and relaxing both the $c$ lattice parameter and the internal coordinates for each case. We have thus explored how the electronic band structure evolves under different degrees of strain and also analyzed the evolution of the electronic structure and magnetic properties as a function of $U$, the on-site Coulomb repulsion.

\section{Ionic picture}

In all these iridates, we have Ir$^{4+}$:d$^5$ cations sitting in an octahedral environment with different degrees of distortions. Because crystal-field splitting is larger than the Hund's rule coupling strength, the ions are in a $t_{2g}^5$ configuration. Due to the large spin-orbit coupling typical in 5d electron electron systems like this, different eigenstates for the single t$_{2g}$ hole may occur.	In the single-ion picture, in the presence of a tetragonal distortion that splits the t$_{2g}$ triplet, two situations may occur depending on the relative strength of spin-orbit coupling with respect to the tetragonal distortion, namely: i) if spin-orbit coupling is just a perturbation, there will be an $xy$ singlet and the $xz/yz$ doublet will be split by spin-orbit coupling into the $xz \ \pm \ i \ yz$ (l$_z$ eigenstates), or ii) if spin-orbit coupling dominates, one should see the splitting caused by spin-orbit coupling into a j$_{eff}$= 3/2 (four-fold degenerate) and 1/2 (two-fold degenerate, higher in energy) states, that become further split when a tetragonal distortion is introduced as second order. The first situation will lead to a spin-half ion with l$_z$ = $\pm$ 1 (L$_z$/S$_z$= 2), and the second situation would produce an Ir$^{4+}$ cation in a state with the following expectation values: l$_z$ = 2/3 and s$_z$ = 1/6 (L$_z$ / S$_z$ = 4). Thus, tracking the ratio of orbital to spin angular momenta, one can describe what particular ionic limit the cation is closer to. Typically, these systems are described as j$_{eff}$ = 1/2 states,\cite{sr2iro4_jeff_1_2} but in principle both situations (and any other intermediate one) can occur, and they will, depending on different variables, such as epitaxial strain, pressure, etc.  The question is what signatures this evolution may show in the band structure, magnetic properties or their strain dependence.


\section{S\MakeLowercase{r}$_2$I\MakeLowercase{r}O$_4$ calculations}

\begin{figure}[t!]
\begin{center}
\includegraphics[width=0.9\columnwidth,draft=false]{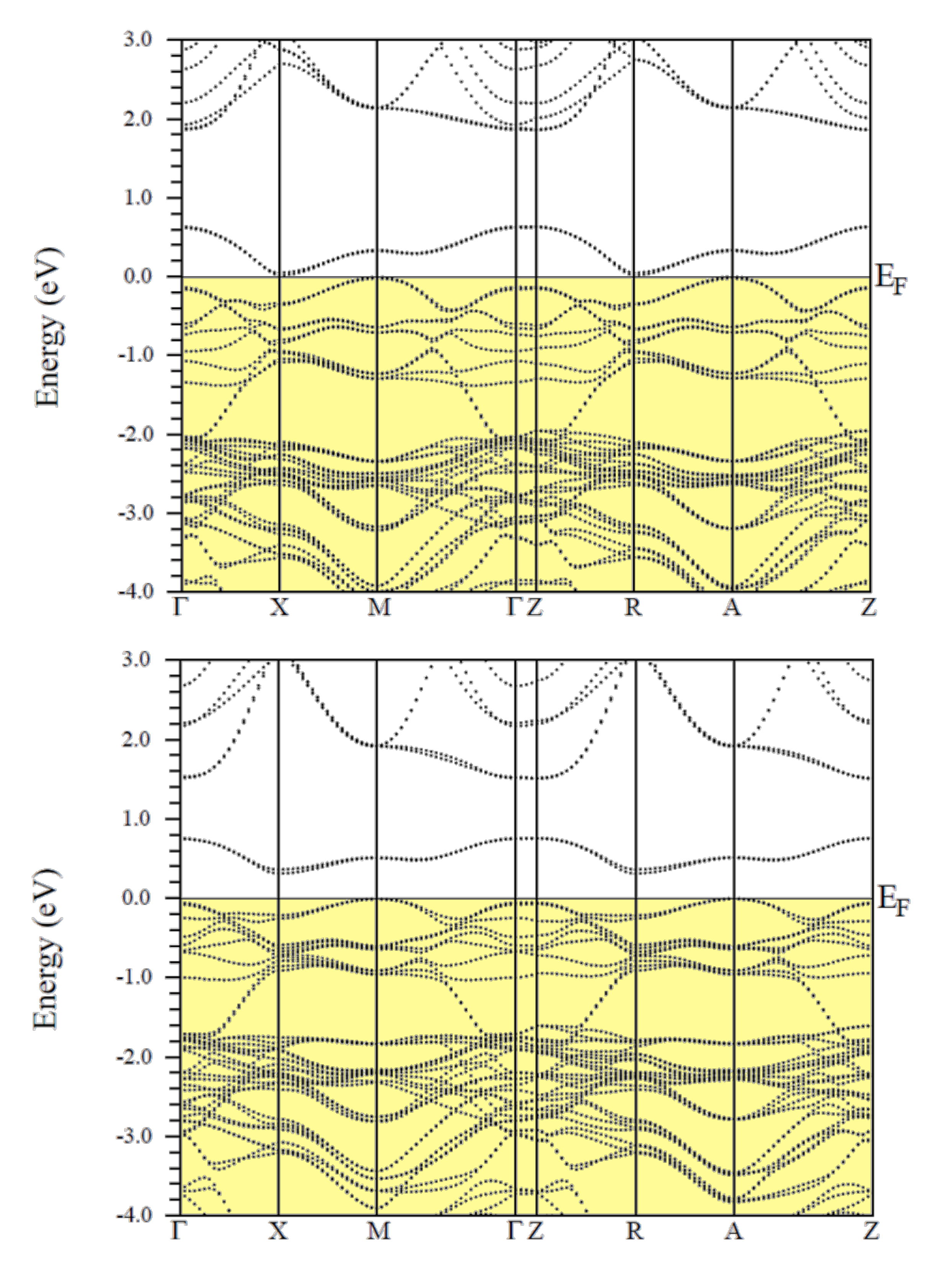}
\caption{(Color online.)  Evolution of the band structure of Sr$_2$IrO$_4$ with strain for U= 3 eV in the ``fully localized limit" utilizing a collinear description for the Ir magnetic moments. Two band structures at different in-plane lattice parameters are presented: on the bottom a larger gap is found for $a$= 3.989 \AA\ (as that of KTaO$_3$) and on top, a smaller gap is obtained for $a$= 3.869 \AA\ (that of LSAT). We can see that the gap becomes reduced as further compressive strain is applied.}\label{sr2iro4_bs}
\end{center}
\end{figure}

As we mentioned above, the analogy to the high-T$_c$ cuprates and the recent hints appearing in the literature indicating possible superconductivity\cite{sr2iro4_sc_1,sr2iro4_sc_2} have drawn considerable attention to this system. The Ir$^{4+}$:d$^5$ cations sit in an elongated octahedral environment, where the t$_{2g}$ levels would be split by a local tetragonal distortion. This structure can be seen in Fig. \ref{sr2iro4}a. The layered structure leads to an in-plane square lattice of Ir atoms (Fig. \ref{sr2iro4}c)  with a singly unoccupied hole, somewhat similar to the situation in cuprates. In this case, the hole is in the t$_{2g}$ manifold (as opposed to the e$_g$ hole in cuprates) and we are dealing with more extended 5d electrons here, compared to the 3d in cuprates. We will see below in the band structure (Fig. \ref{sr2iro4_bs}) that the O p levels are away from the Fermi level in these iridates, which could provide a difference with cuprates. This also occurs for the layered low-valence nickelates,\cite{la4ni3o8_original,la4ni3o8_vpardo,la4ni3o8_sst,layered_nickelates_mit,la4ni3o8_goodenough} which were also suggested as possible candidates for cuprate-like physics because they show similar electron count on the Ni square lattice. AF coupling (with a slight canting) of the Ir moments is shown in Fig. \ref{sr2iro4}c. Nearest-neighbor AF exchange also resembles the in-plane checkerboard pattern typical in cuprates.

In order to describe the electronic structure of the system, let us first try to draw some light on the controversy of how the band structure evolves with tensile strain. For this sake, we have performed ab initio calculations fixing the in-plane lattice parameter to various typical $a$ values corresponding to the above mentioned usual substrates utilized to grow thin films on. For each of those, we have optimized the atomic positions and the $c$ out-of-plane lattice parameter.

\begin{figure}[t!]
\begin{center}
\includegraphics[width=0.9\columnwidth,draft=false]{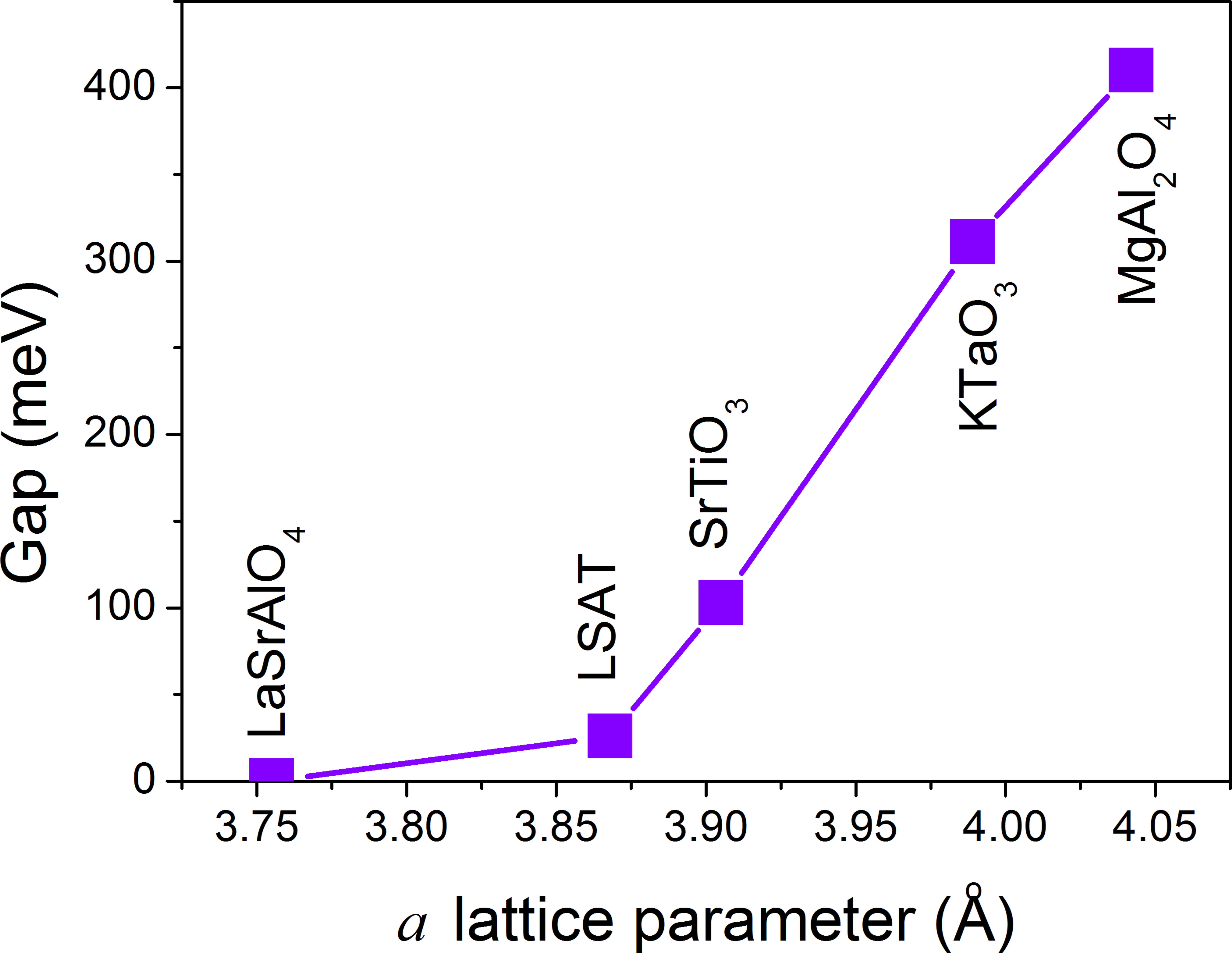}
\caption{(Color online.)  Evolution of the band gap of Sr$_2$IrO$_4$ as a function of the in-plane lattice parameter (tensile [compressive] strain to the right [left]) for U= 3 eV in the ``fully localized limit" utilizing a collinear description for the Ir magnetic moments. We see that tensile strain leads to a larger gap by reducing the in-plane hopping. The trends are consistent for other U values but the actual metal-insulator transition is largely dependent on the U chosen.}\label{sr2iro4_gap}
\end{center}
\end{figure}

In principle, one could reason that in the case of Sr$_2$IrO$_4$, with a layered structure, increasing the in-plane lattice parameter would tend to reduce the in-plane Ir-Ir hopping and the corresponding reduction in $c$ brought about by tensile strain will not produce any additional band broadening. The latter makes sense because of the layered structure, as one can see in Fig. \ref{sr2iro4}, where negligible direct Ir-Ir hopping along the c axis is anticipated. Thus, the expected situation from an electronic structure point of view would be an increased metallicity as it is compressed in the plane. This would be in agreement with XAS measurements in Ref. \onlinecite{sr2iro4_gap_vs_strain} and in disagreement with gap estimates as a function of strain in Ref. \onlinecite{sr2iro4_gap_vs_strain_2}, that show the opposite trend.

Figure \ref{sr2iro4_gap} shows the results of our ab initio calculations for the evolution of the gap with strain for a particular U value (U= 3 eV). The trends shown would be consistent for other U values, but the particular gaps obtained, and also the point for the metal-insulator transition would differ (the evolution with U will be analyzed in further detail below in Section \ref{u_evol}). We chose this (probably) large U value with a broad insulating region for the sake of illustrating more clearly the evolution with strain.
We can see in Fig. \ref{sr2iro4_gap} how the gap becomes larger as in-plane tensile strain is applied, as one would expect from the simplistic electronic structure arguments we just explained. As the material is compressed in the plane, larger bandwidths are obtained and eventually the system becomes metallic at short enough Ir-Ir in-plane distances.

We show in Fig. \ref{sr2iro4_bs} the basic band structure of the system for two different substrates to illustrate further the way the gap gets smaller as in-plane compressive strain is applied. Again, calculations are presented for a large U of 3 eV so that a gap opens up and its strain dependence can be analyzed more clearly. We see that the band closing comes about without changing the main features of the band structure or the dispersions, it is very much a rigid band shift (plus some additional band broadening).

We proceed to describe the electronic structure of Sr$_2$IrO$_4$ in more detail. As explained above, the Ir$^{4+}$:d$^5$ cation in a (distorted) octahedral environment will have one hole in the t$_{2g}$ manifold. Our ab initio calculations show that when the LDA+U method is used for any finite U, an in-plane AF ordering is obtained, and we have used this configuration for all the results analyzed throughout the paper. The relative magnitude of the spin and orbital moments depends both on strain and the U value (the evolution with U will be analyzed in detail in Section \ref{u_evol}). The effect of U is rather simple, increasing the on-site Coulomb repulsion leads to larger moments, both orbital and spin components. However, the effect of strain is not easy to predict a priori.

\begin{figure}[t!]
\begin{center}
\includegraphics[width=0.9\columnwidth,draft=false]{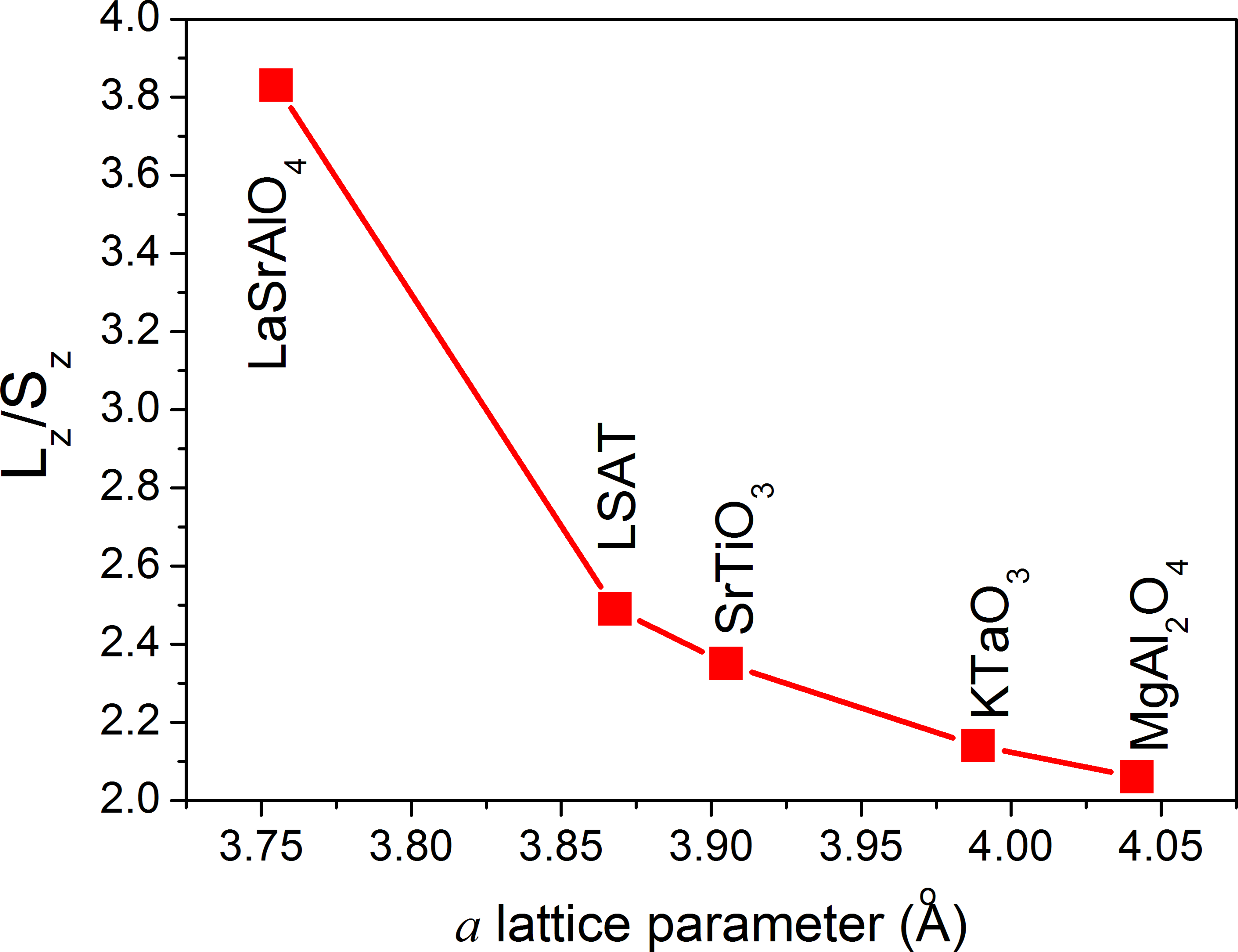}
\caption{(Color online.)  Evolution of the L$_z$/S$_z$ ratio of Sr$_2$IrO$_4$ with strain for U= 3 eV in the ``fully localized limit" utilizing a collinear description of the magnetic moments. We can see that the values are always close to 2.0, being closer to 2.0 for the more insulating solutions in the tensile strain limit.}\label{sr2iro4_lz_sz}
\end{center}
\end{figure}

We can observe in Fig. \ref{sr2iro4_lz_sz} the evolution of the L$_z$/S$_z$ ratio as a function of strain for a fixed value of U= 3 eV, the same one we used for analyzing trends of the band gap with strain. Let us recall that a value of this ratio closer to 2.0 corresponds to a hole in an $xz\ \pm\ i\ yz$ orbital, and values closer to 4.0 would in principle correspond to the so-called j$_{eff}$=1/2 solution. We observe that all values are closer to 2.0 except at the metallic limit, where the moments have almost vanished ($a$ of LaSrAlO$_4$). In that case, the ratio is very large but it does not indicate a j$_{eff}$=1/2 solution was encountered. The tendency shows that the more insulating phases tend to stabilize a ratio closer to 2.0.

Consensus in the literature describes almost unanimously this material as being a realization of a j$_{eff}$= 1/2, with a certain degree of mixing with the lower-lying j$_{eff}$= 3/2 states. However, we have just seen that within a collinear formalism using \textsc{wien2k} (which is the methodology utilized for all the calculations presented up to this point), no solution that resembles a j$_{eff}$= 1/2 state, in terms of L$_z$/S$_z$ ratio (or the separate values of L$_z$ and S$_z$ expected for that simplistic single-ion solution) can be found. Changing U, the initial conditions for magnetism, or even the LDA+U flavor (around the mean field\cite{amf} was also tested) helps modifying slightly the actual values but not the main conclusions that the system is closer to having one hole in an $xz \ \pm \ i \ yz$ orbital. One could expect that for a spin-polarized system, the spin-up and down channels can be substantially separated in energy so that spin-orbit coupling can barely mix them. Yet, the purely ionic $j_{eff}$=1/2 solution has an expectation value of $S_z$ of only 1/6, maybe not enough to induce a large spin-up / spin-down Hund's related splitting. However, this seems to be the case according to the calculations.

It is worth noting again, as can be seen slightly in Fig. \ref{sr2iro4_bs}, that changing the L$_z$/S$_z$ ratio does not seem to modify the band structure. In the cases presented there, the application of strain basically leads to a rigid shift in the bands (plus some band broadenings due to increased in-plane hoppings), but the evolution to a larger ratio at more in-plane compressed lattices does not induce a change in the band dispersions. However, we will show now that a change in the band character of the t$_{2g}$ hole does occur even if the bands look unaffected. We will see that the L$_z$/S$_z$= 2.0 limit presents a hole that is fully composed of an $xz \pm i\ yz$ orbital, and going towards a larger ratio modifies the occupancies mixing different orbital contributions.

\begin{figure}[t!]
\begin{center}
\includegraphics[width=0.8\columnwidth,draft=false]{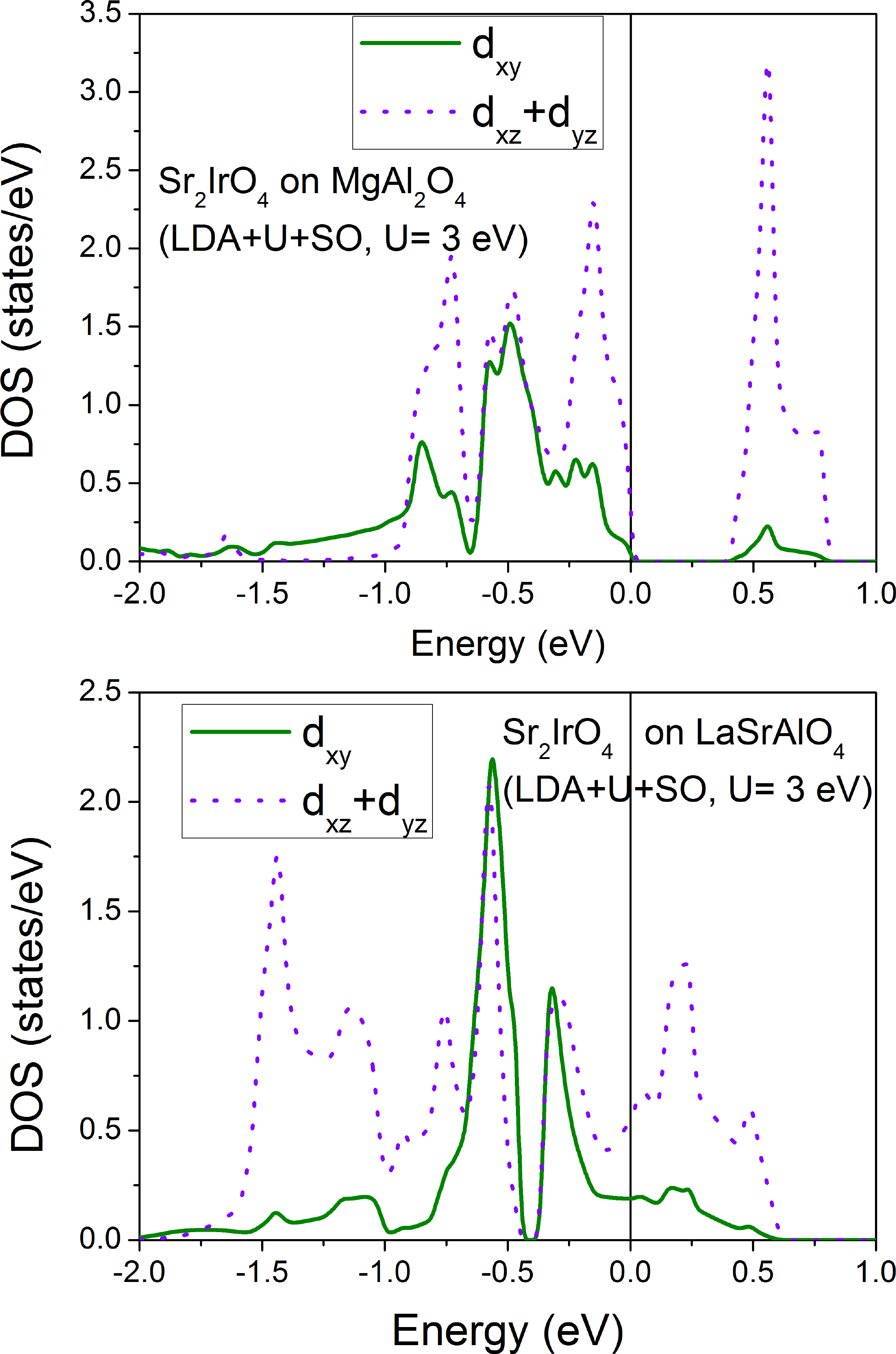}
\caption{(Color online.) Projected density of states for Sr$_2$IrO$_4$ in the limit of L$_z$/S$_z$= 2.0 (top, with $a$ of MgAl$_2$O$_4$) where the hole is clearly in an $xz$/$yz$ combination with no $xy$ character, and an increased ratio on the bottom (with $a$ of LaSrAlO$_4$) where a larger $xy$ character of the t$_{2g}$ hole is observed. These are collinear calculations with U= 3 eV within the ``fully localized limit".}\label{sr2iro4_dos}
\end{center}
\end{figure}

To analyze this change, we present in Fig. \ref{sr2iro4_dos} two DOS curves at different in-plane strains. Even though we saw before that moving towards a larger in-plane strain barely changes the band structure (only a rigid shift), the evolution towards a more metallic state induces a larger $xy$ mixing in the t$_{2g}$ hole. However, the j$_{eff}$=1/2 solution is not obtained, there is always much larger $xz/yz$ contributions, which is incompatible with having a j$_{eff}$=1/2 state. One could think that this has to do with the huge tetragonal distortion of the octahedral environment.\cite{sr2iro4_pressure} However, at large tensile strains, such as the case of setting $a$ to that of MgAl$_2$O$_4$, the in-plane Ir-O distance becomes equal (only 0.3 \% difference) to the out-of-plane metal-anion bond length. Thus, the reason behind this unmixing does not have to do with the on-site energies of the different t$_{2g}$ levels that are split by the oxygen crystal field. Instead, it is the relative strength of spin-orbit coupling, that acts as a perturbation to the crystal field and splits the $xz/yz$ orbitals forming l$_z$ = $\pm$ 1 eigenstates,\cite{sto_svo} at least in a collinear calculation, since
up to this point all the calculations presented were carried out within a collinear formalism to treat magnetism. It is worth noting that within this collinear scheme within {\sc wien2k}, the L$_z$/S$_z$ ratio is very much independent of the magnetization direction, whether it lies in the plane or out of the plane. The moments (both orbital and spin components) become larger (by up to 30\%) when the magnetization is out of the plane. We will describe in more detail the magnetization direction within a non-collinear description, which yields an orientation of the magnetic moments consistent with experimental observations.

\begin{figure}[t!]
\begin{center}
\includegraphics[width=1.0\columnwidth,draft=false]{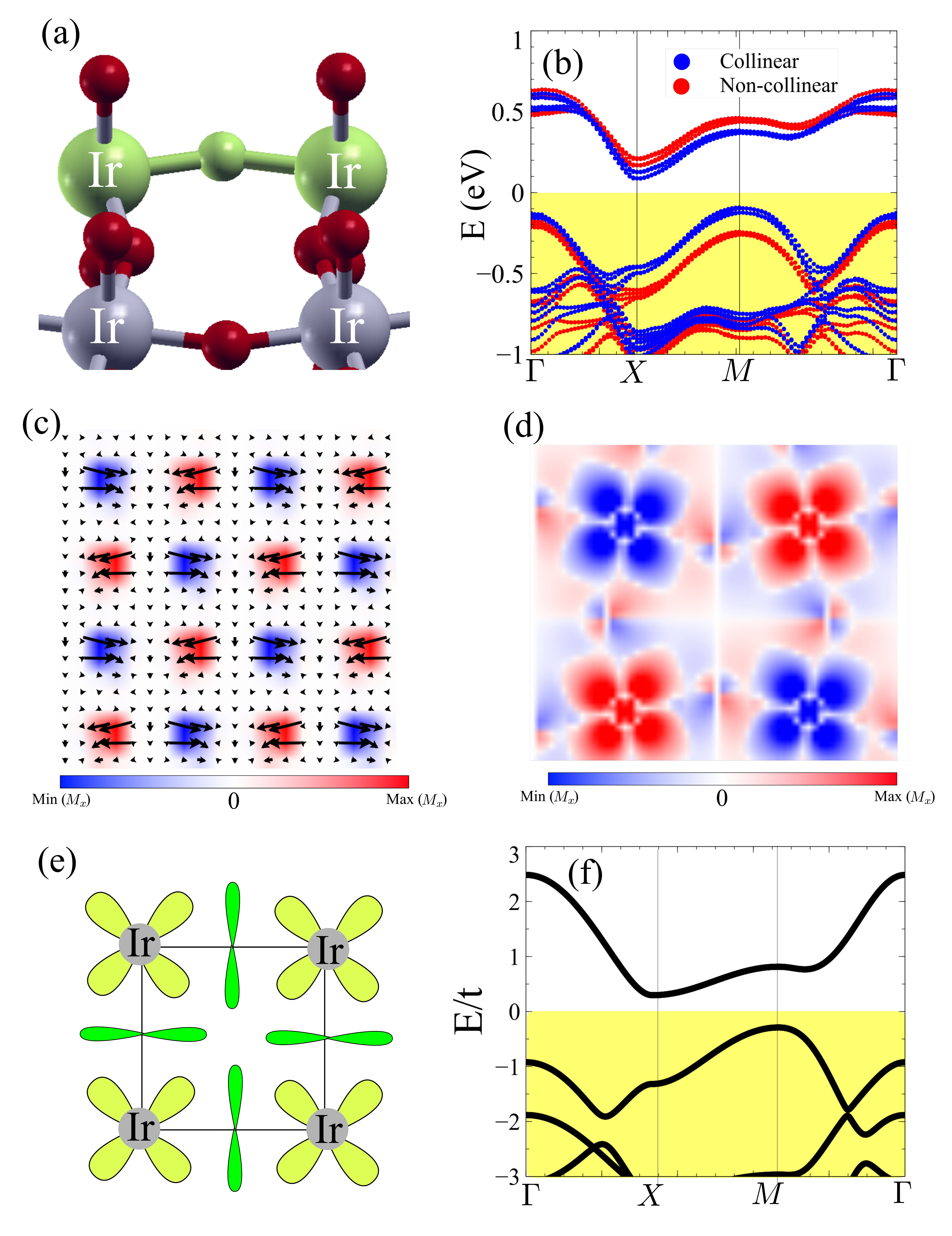}
\caption{(Color online.) (a) Scheme of the Ir-O-Ir
distortion in Sr$_2$IrO$_4$, which gives rise to weak ferromagnetism.
(b) Comparison of band structures calculated with a collinear
and non-collinear scheme. (c) Vector plot in the Ir plane, showing 
the local magnetization for the different Ir atoms as obtained
in the DFT non-collinear calculation, showing
the canted antiferromagnetic state. Panel (d) shows a zoom close to the Ir atoms,
showing a magnetization contour plot shaped as a $t_{2g}$ orbital as well as a small
contribution from the oxygen atoms.
(e) Sketch of a minimal model to describe the electronic structure
of Sr$_2$IrO$_4$, by taking into account nearest neighbors hoppings
(only one $t_{2g}$ orbital is shown in each metal atom and the corresponding p$_{\pi}$ orbital in the neighboring oxygen). Red-blue orbitals
are $d_{xy}$ orbitals, and green is an intermediate oxygen orbital.
(f) Electronic band structure of the tight binding model, obtained
by means of a Hartree Fock calculation of the Hubbard model.
The parameters used are $\lambda_{SOC} = t $ and $U=8t$. No Ir-O-Ir angle canting
nor octahedral distortions have been taken into account in the model.} 
\label{sr2elk}
\end{center}
\end{figure}

Calculations were also carried out with the {\sc elk} code, both in the
collinear and non-collinear formalism. Within both schemes, we
take the initial magnetization to point in-plane along the Ir-Ir direction. As shown in Fig. \ref{sr2elk}b, the band structures within
a collinear and a non-collinear scheme for magnetism are very similar, the same kind of dispersions and only some rigid shifts occur between both methodologies.
When looking at the local magnetic order, we observe that
the expectation values of $\vec L$ and $\vec S$ are not
completely parallel, the ground state is a slightly canted antiferromagnet, produced due to the deviation of the Ir-O-Ir angle away from 180$^{\circ}$ 
(see the zoomed region in Fig. \ref{sr2elk}a).

A small net moment is obtained in each Ir atom, apart from the
AF component, yielding a net total angular momentum (spin plus orbital) $J_{FE} = 0.042$.
This arises due to the non-collinearity
between the local magnetism of neighboring
Ir atoms (see Fig. \ref{sr2elk}c,d).
Within the
non-collinear DFT calculations, the
angle between the magnetic moments in neighboring atoms
is 151$^{\circ}$ (180$^{\circ}$ would be collinear antiferomagnetism).
Moreover, the angle between L and S is $\alpha = 8.1$$^{\circ}$.
Therefore, the small net ferromagnetic (FM) component that
arises due to the Ir-O-Ir angle,
is expected to yield a weak FM
signal in Sr$_2$IrO$_4$. This is in qualitative agreement with experimental observations\cite{sr2iro4_mu_eff} showing an ordered moment of about 0.14 $\mu_B$. In our case, the value can be smaller because only the value inside the muffin-tin spheres is considered and the 5d electrons that cause it are very spread into the interstitial region.
The ratios between orbital and spin components yield $L/S = 5.4$ and $L_z/S_z = 5.6$ for this non-collinear scheme.
These are even larger than the
purely ionic description of the j$_{eff}$= 1/2 state, and are certainly very far from being a representation of a hole in an $xz$ $\pm$ $i$ $yz$ state. However, this is not reflected in the band structure, which barely changes when this ratio becomes so drastically modified.
Remarkably, the band structures of the magnetic non-collinear solution
are very similar to those obtained with a collinear scheme,
as shown in Fig. \ref{sr2elk}b.
For all these calculations, we have used a value of U= 2.7 eV, nevertheless
in Section 
\ref{u_evol} we will discuss the dependence of
the ratio $L_z/S_z$ with
the on-site Coulomb
repulsion U.

Leaving apart the small canting of the Ir magnetic moments,
 the main features of the electronic structure can be captured with a simple
tight binding model for the $t_{2g}$ orbitals in a square lattice of the form
\begin{equation}
H = H_{NN}+H_{SOC}+H_U
\end{equation}
where $H_{NN}$ is the nearest neighbors tight-binding hopping,
$H_{SOC}$ the SOC projected onto the $t_{2g}$ manifold and $H_U$
the local electron-electron interaction which
will give rise to magnetic order.
Taking local $t_{2g}$ orbitals (labeled by $v,w$) in each
Ir atom (labeled by $i,j$), we only considered as
non-vanishing hopping terms those
involving indirect hopping through the
oxygen atom situated between them. This situation is sketched in Fig. \ref{sr2elk}c.
\begin{equation}
H_{NN} =\sum_{v,i,w,j} \nu_{v,i,w,j} c^\dagger_{v,i} c_{w,j}
\end{equation}
where $\nu_{v,i,w,j}$ can be easily obtained by symmetry considerations.

We include SOC by calculating
the matrix elements of the SOC operator for the $t_{2g}$ orbitals.
\begin{equation}
H_{SOC} =\Lambda_{SOC} \vec L \cdot \vec S
\end{equation}

 With such a nearest neighbor
tight binding model, we have also introduced electronic interactions
in a minimal way with an intraorbital Hubbard model
\begin{equation}
H_U = \sum_{i,w}Uc^\dagger_{i,w,\uparrow} c_{i,w,\uparrow} 
c^\dagger_{i,w,\downarrow} c_{i,w,\downarrow}
\end{equation}  
 and we solved the system with a Hartree Fock mean field approximation.

With those ingredients, the following features are obtained. The ground state of the system at $t_{2g}^5$ filling is AF in agreement with the DFT calculations.
Focusing on a particular Ir atom,
the expectation value of the magnetic moment
changes sign between the different
 $t_{2g}$ orbitals, but giving a net contribution,
in agreement with the $j_{eff}$ = 1/2 picture. Let us recall that those $j_{eff}$ eigenstates look like the following in terms of the standard t$_{2g}$ orbitals: $\left| j= 1/2 ; j_z = \pm 1/2 \right> = \left( \left| yz, \pm \sigma \right> \mp i \left| xz, \pm \sigma \right> \mp \left| xy, \mp \sigma \right> \right)/ \sqrt{3}$, where $\sigma$ describes the spin.
 Magnetic order introduces
the same kind of band splitting which ultimately gives
rise to the insulating behavior. And finally, the band
structures (Fig. \ref{sr2elk}f) resemble in dispersions and locations of maximum/minima the ones obtained
in our ab initio calculations. Nevertheless, we have not implemented the Ir-O-Ir canting angle into this simple tight binding
model, and therefore it
is not able to capture the weak ferromagnetism of Sr$_2$IrO$_4$. Yet, the band structure obtained matches pretty well the main features of the electronic structure (see Fig. \ref{sr2elk}f vs b), and naturally produces a hole in a state close to the $j_{eff}$ = 1/2 ionic description.

\section{S\MakeLowercase{r}I\MakeLowercase{r}O$_3$ calculations.}

We have also run calculations in the structurally simpler SrIrO$_3$, whose low-temperature structure is an orthorhombic distorted perovskite.\cite{sriro3_struct} For imposing different in-plane strains simulating an epitaxial growth on a cubic substrate, we have used a tetragonal perovskite structure, allowing for octahedral rotations and relaxing these using the GGA-PBE exchange correlation potential. Again, we have imposed different $a$ lattice parameters as chosen above for the case of Sr$_2$IrO$_4$, optimizing the atomic positions and the off-plane lattice constant.

Experimental evidences from thin films grown on top of different substrates show that compressive strain applied to SrIrO$_3$ increases the resistivity.\cite{sriro3_mit} The situation in the perovskite-based system is different to the layered Sr$_2$IrO$_4$. Here, compressing in the plane implies (due to the approximate volume conservation) elongating in the off-plane direction, so it is not clear how this can affect the overall band structure, and in particular the opening/closing of a band gap around the Fermi level. On the one hand, compressive strain increases the in-plane hoppings but it reduces the out-of-plane ones. The situation is more complex here.

\begin{figure}[t!]
\begin{center}
\includegraphics[width=0.9\columnwidth,draft=false]{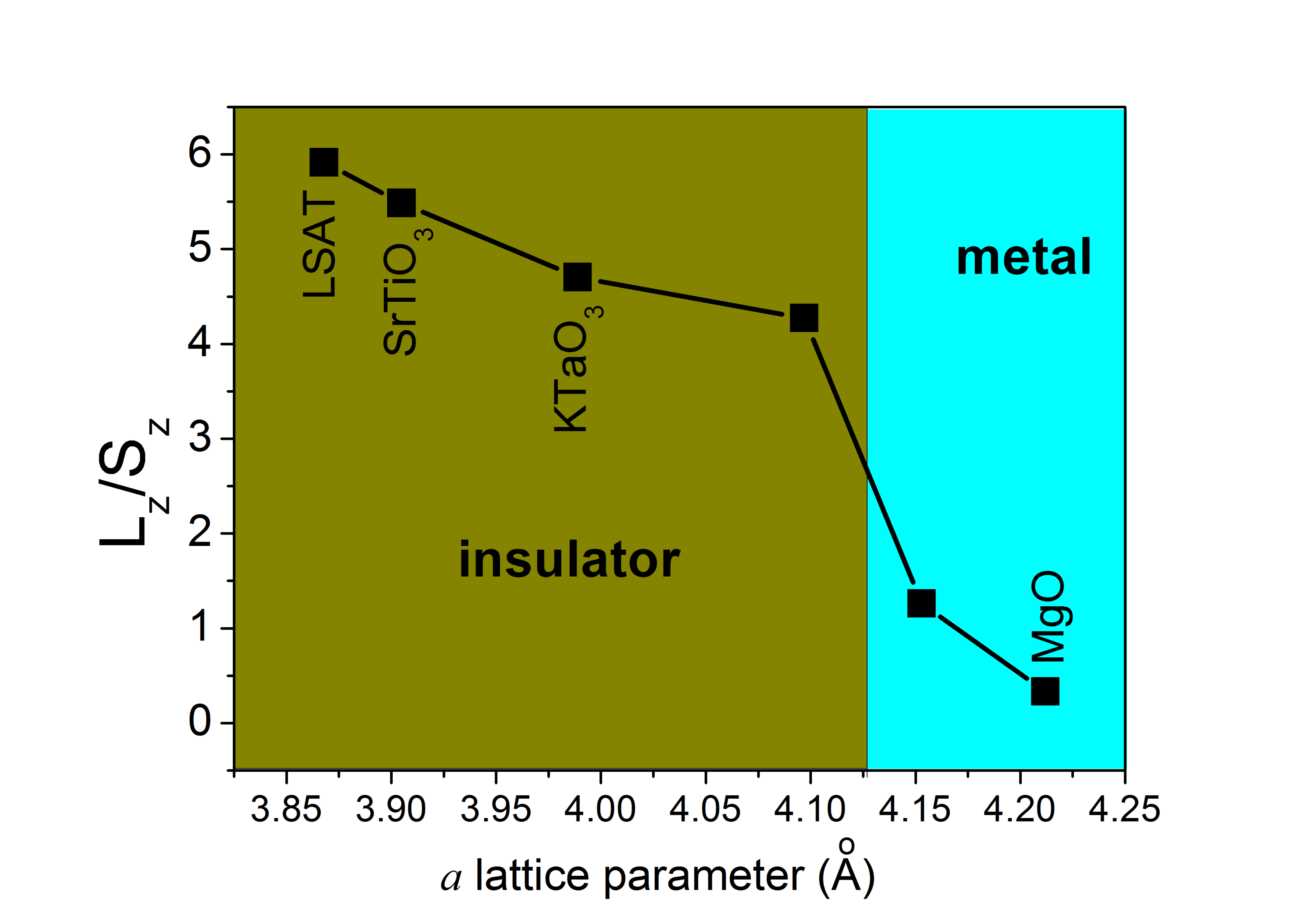}
\caption{(Color online.) Evolution with in-plane strain of the L$_z$/S$_z$ ratio for a non-collinear calculation in SrIrO$_3$. Values on the order of 4.0 and higher are obtained through a wide range of in-plane strains, particularly within the insulating regime (whose extent depends on the particular U value). Results with U= 2.7 eV in the Yukawa-screened version are presented here, the trend with strain being systematic for a wide range of U values. }\label{lz_sz_elk}
\end{center}
\end{figure}

To analyze this compound, we have carried out calculations using a fully relativistic non-collinear scheme using the {\sc elk} code, setting up an AF coupling between Ir nearest neighbors (so-called G-type AF ordering) and introducing on-site Coulomb repulsion via the LDA+U method. By imposing different $a$ lattice parameters and optimizing both the c lattice parameter and the internal coordinates, we can obtain the evolution of the band gap with strain. Calculations show that the band gap becomes smaller as in-plane tensile strain is applied, in agreement with experiments. In particular, if a U= 2.7 eV is used within the Yukawa-screened formalism (see Fig. \ref{lz_sz_elk}), SrIrO$_3$ becomes metallic for $a$ $\sim$ 4.12 \AA, approximately, but this crossing point depends heavily on the U value chosen. This trend is in principle consistent with the experimental evidences described above. Elongating in the plane produces a larger compression along the c-axis in order to retain an optimal volume, and that small Ir-Ir off-plane distance eventually leads to a metallic bonding. We have not tested the situation with comparable Ir-Ir distances in the plane, but we expect that for a sufficient compressive strain, the gap will start to decrease as well.

Figure \ref{lz_sz_elk} shows the evolution of the L$_z$/S$_z$ ratio obtained within a non-collinear scheme. We observe that the values are significantly higher than those we obtained within a collinear scheme using {\sc wien2k} for Sr$_2$IrO$_4$, and more consistent with those values yielded by {\sc Elk} for Sr$_2$IrO$_4$ in a non-collinear calculation and with the hole description provided by the tight-binding model. These values resemble more the j$_{eff}$= 1/2 solution often quoted in literature to explain the electronic structure of this series of materials, and in some cases the ratio even exceeds that of the single-ion solution. When the metallic region is reached, the ratio drops drastically. Again, the transition point between insulating and metallic phase will depend strongly on the value of U chosen. We picked U= 2.7 eV (in the Yukawa-screened flavor) since it allows us to describe the trends more clearly due to the enhanced (probably unrealistic) gaps, and these trends with strain are consistent in a broad range of U values.


\begin{figure}[t!]
\begin{center}
\includegraphics[width=0.7\columnwidth,draft=false,angle=-90]{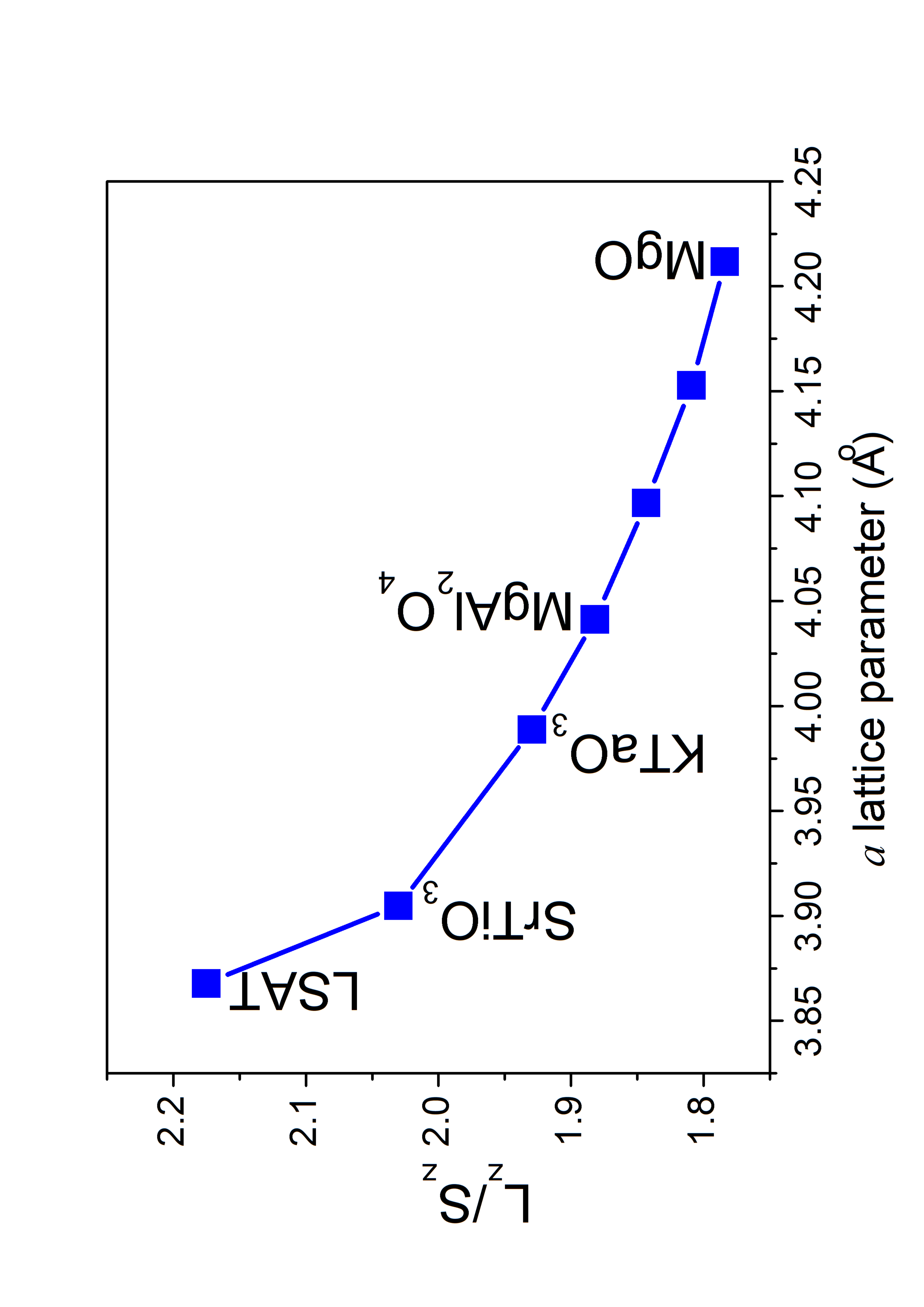}
\caption{(Color online.) Evolution with in-plane strain of the L$_z$/S$_z$ ratio for a collinear calculation in SrIrO$_3$. The evolution with strain is also towards a smaller ratio at larger tensile strain, but the values are in a completely different limit. These particular values were obtained for U= 3 eV in the ``fully localized limit". We see that, like in Sr$_2$IrO$_4$, a solution with ratio close to 2.0 is the most stable one.}\label{lz_sz_col}
\end{center}
\end{figure}

Let us now try to draw some comparisons with our
previous calculations in Sr$_2$IrO$_4$ using a collinear
scheme as implemented in {\sc wien2k}
now for SrIrO$_3$. The results are quite different, resembling more the picture described above for the layered compound. We can see them summarized in Fig. \ref{lz_sz_col}. We observe that again the values of the L$_z$/S$_z$ ratio are closer to 2.0, just like we saw for Sr$_2$IrO$_4$. We even tried to introduce the density matrices of a solution as obtained in the non-collinear scheme in the collinear calculation, but it never converges without evolving towards the solution close to 2.0. This is quite unique, since the LDA+U method is usually capable of converging many different solutions, as long as the appropriate density matrices are used at the initial step. The evolution with strain towards smaller ratios is consistent in both schemes, but on a completely different scale of values.


\begin{figure}[t!]
\begin{center}
\includegraphics[width=1.05\columnwidth,draft=false]{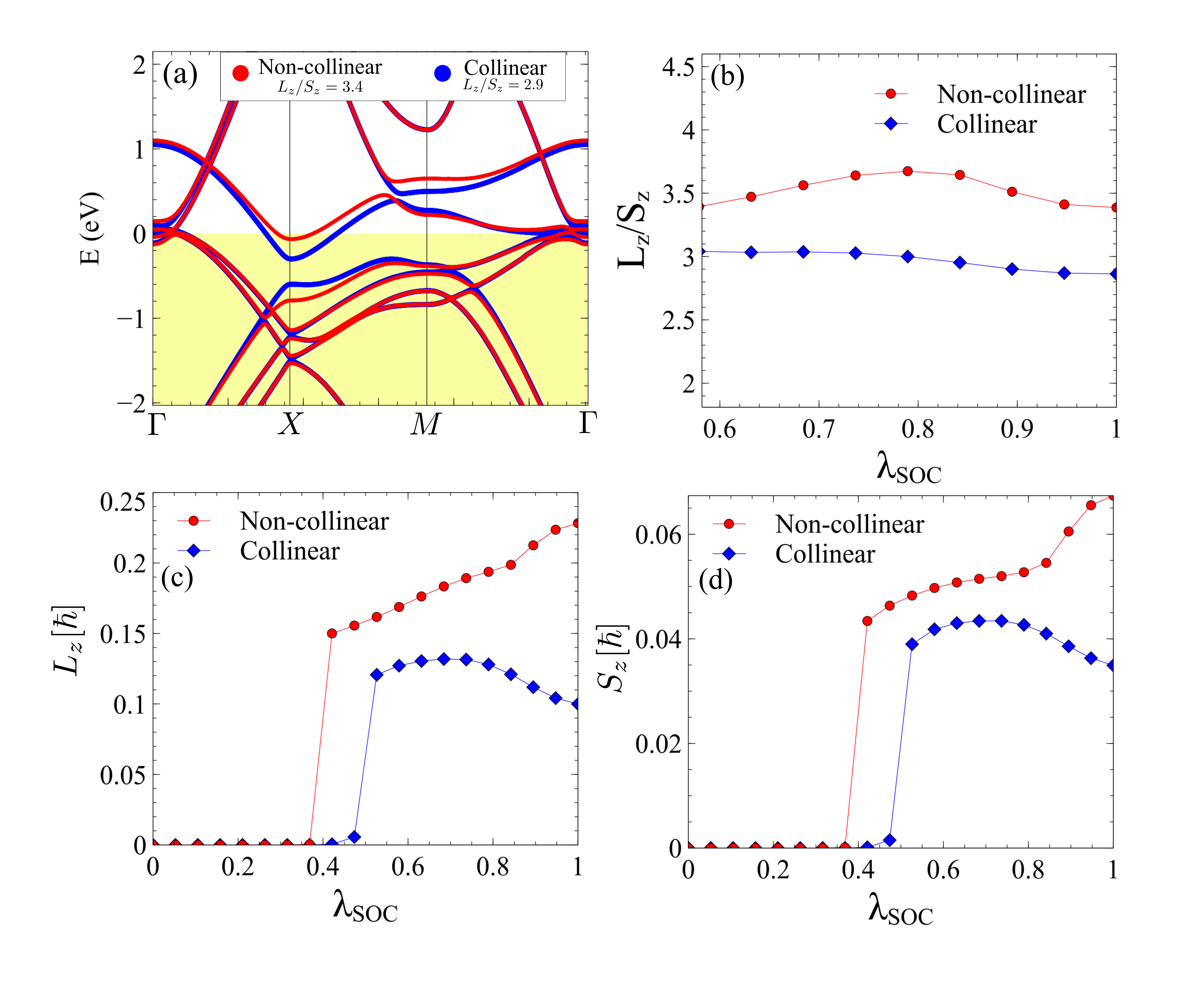}
\caption{(Color online.) 
Calculations for tetragonal SrIrO$_3$ with 1.5\% uniaxial compressive deformation.
(a) Comparison
of the band structures with AF order (along the uniaxial axis) for both a collinear and a non-collinear
scheme. (b) Evolution with
spin-orbit coupling strength of the L$_z$/S$_z$ ratio for
a non-collinear and a collinear calculation.
Evolution with SOC strength of $L_z$ (c) and $S_z$ (d), showing that
switching on SOC triggers the magnetic order in the system. 
Noncollinear calculations yield 
larger $L_z$ and $S_z$ values, and also an increased
$L_z/S_z$ ratio.
Calculations are performed with $U=2.7$ eV in the Yukawa-screened scheme.}
\label{toy_model}
\end{center}
\end{figure}

\section{Evolution with SOC}

Trying to dig a bit more into the origin of this behavior,
in this section we will study the dependence
of the magnetic/orbital order
with SOC strength.  We have set up
a tetragonal  
$\text{SrIrO}_3$ structure
with 1.5\% tetragonal compressive uniaxial distortion
in an AF order to be used as 
toy model.
We stress that the incoming discussion
is about a minimal model to understand the behavior
of $j_{eff}$ within the collinear and non-collinear schemes. We performed
LDA+U calculations with $U=2.7$ eV in the Yukawa-screened scheme. 
In order to understand the effect of SOC,
we have tuned the strength of the SOC from the non-relativistic
limit $\lambda_{SOC}=0$ to its real value $\lambda_{SOC}=1$,
where $\lambda_{SOC}$ is a multiplicative constant which scales the 
relativistic part of the Hamiltonian in the valence electrons
\begin{equation}
H = H_{non-relativistic} + \lambda_{SOC} H_{relativistic}
\end{equation}
and we follow the evolution of the $L_z/S_z$ ratio
and the order parameters $L_z$ and $S_z$. We perform such study
for both a fully non-collinear scheme and a 
collinear scheme in the direction of the AF order. 
With this toy model, we obtained
the L$_z$/S$_z$ ratio as a function of
spin-orbit coupling strength that we
can see in Fig. \ref{toy_model} b. This
shows an important difference in ratios
only because of setting up a
non-collinear calculation. 
As
in the previous results, the collinear calculation yields smaller
 ratios throughout the different
SOC strengths considered.

Importantly, by following the evolution of $L_z$ 
(Fig. \ref{toy_model}c) and $S_z$ (Fig. \ref{toy_model}d),
it is clearly observed
that SOC triggers the magnetic order in the system. This can
be understood as follows: in the large SOC limit, the Fermi level
lies in the middle of the $j_{eff}=1/2$ manifold,
which has a smaller bandwidth than the full non-relativistic
$t_{2g}$ manifold  so that the system
would have a large DOS at the Fermi level. In the intermediate regime,
by switching on SOC,
the system undergoes a transition from the non-relativistic
$t_{2g}$ limit to the relativistic $j_{eff}=1/2$ state, which
turns the systems vulnearable towards symmetry breaking due
to the decreasing bandwidths near the Fermi energy, in particular towards magnetism.

Furthermore,
non-collinear calculations yield
important increases of $L_z$ and $S_z$
with respect to the collinear case.
This difference also implies additional (minor) splittings
in the band structure, as shown in Fig. \ref{toy_model}a.
It is worth to note that in the present system, the Ir-O-Ir
bond is linear, so that this difference does not arise due
to the canting of moments between different Ir atoms,
Therefore, the difference in ratios between collinear
and non-collinear schemes is related with the internal non-collinear
nature of the
magnetism within each atom.
The present
phenomenology suggests that non-collinearity
might be of critical importance when exploring $j_{eff}$
order.

\section{Evolution with U}\label{u_evol}

\begin{figure}[h]
\begin{center}
\includegraphics[width=1.0\columnwidth,draft=false]{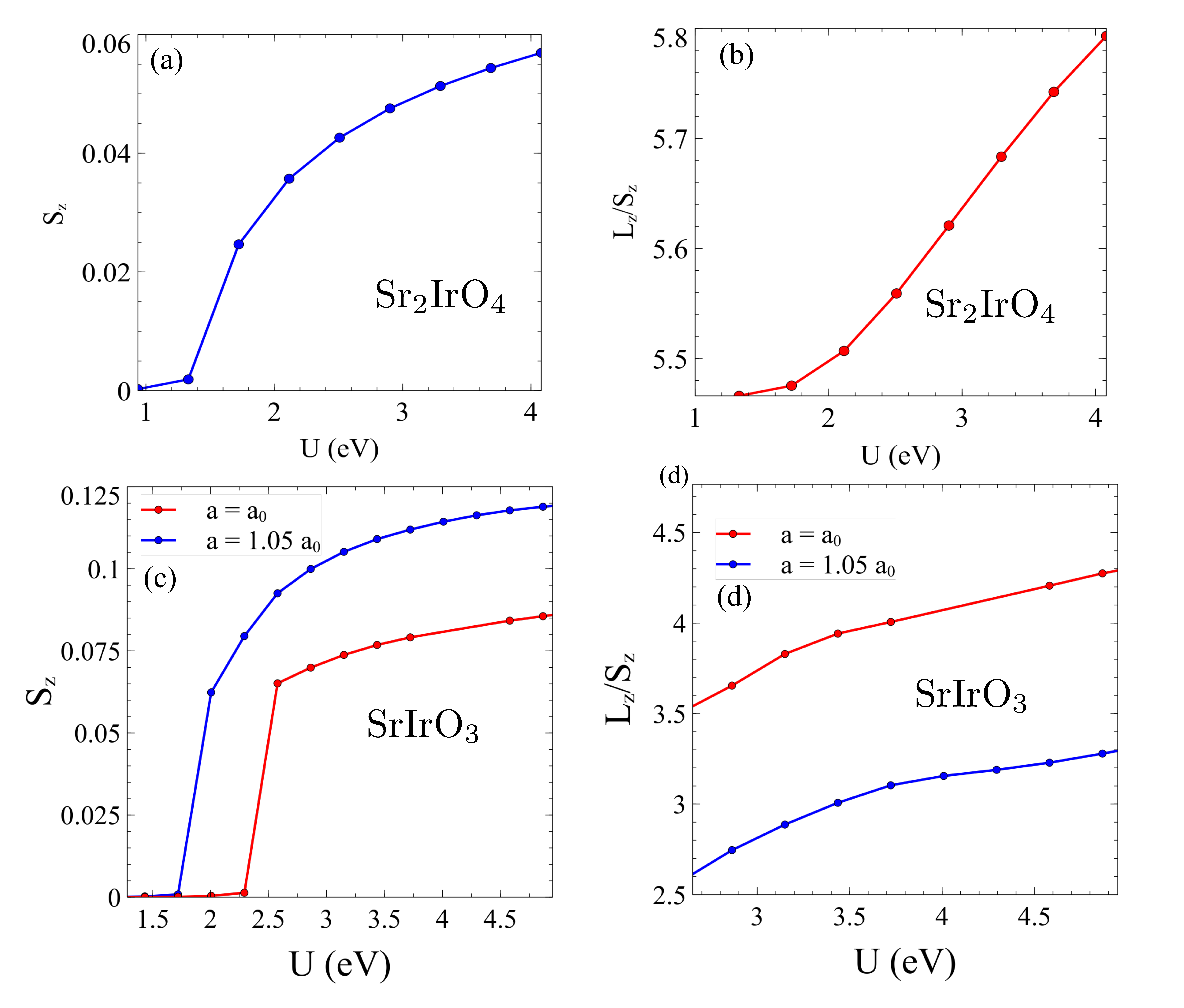}
\caption{Evolution of the spin moment in the $z$ direction
for Sr$_2$IrO$_4$ (a) and SrIrO$_3$ (c) as a function of the on-site Coulomb repulsion.
The panels on the right show the evolution of the ratio $L_z/S_z$ for Sr$_2$IrO$_4$ (b) and SrIrO$_3$ (d).
Both systems show an increasing ratio with increasing on-site Coulomb repulsion.
All the calculations where performed within a non-collinear formalism. }
\label{evolu}
\end{center}
\end{figure}

In this section we will discuss the effect of the electron-electron interactions
inside the d-manifold, as taken into account in an LDA+U scheme. In order to complete the previous discussion, we have performed fully non-collinear calculations
with the { \sc elk} package. For a systematic study of the dependence with U of the electronic structure and magnetic properties, we have introduced U by means of the Yukawa scheme
and fixing it to different values.  The main results
are summarized in Fig. \ref{evolu}.

For Sr$_2$IrO$_4$, magnetic order starts
to show up from $U=1.5$ eV, increasing rapidly with $U$, until reaching
a value of $S=0.06$ for $U=4$ eV (which does not seem to be fully saturated). In comparison, the $L_z/S_z$ ratio
undergoes only small changes upon increasing U, showing a value around
$L_z/S_z=5.7$, above the expected value of 4 for the ionic limit $j_{eff}=1/2.$

For SrIrO$_3$, however, (Figs. \ref{evolu}c and \ref{evolu}d) we can observe a critical value of $U$ from which magnetism
sets in. It is worth noting that even when the system starts to become
magnetic, it retains its metallicity until a larger U value is reached.
For a lattice constant $a_0 = 3.95$ \AA\ the critical U for the onset of magnetism is $U_C = 2.5$ eV, whereas
upon an artificial volumetric expansion of $5\%$ in the lattice constant,
the critical U lowers to $U_C = 1.7$ eV. Even though such
volumetric expansion is not fully realistic, it provides an insight on the
effect of the Ir-Ir distance in the critical $U$. Regarding the $L_z/S_z$
ratio, it is observed that the ratio increases with U, but decreases
upon increasing the lattice constant, as we explained in detail above when dealing with strain effects.

\section{Concluding remarks.}

To summarize, in this paper we present ab initio calculations on Ir$^{4+}$-based oxides, which provide an electronic structure understanding of the evolution of its electronic structure, magnetic and transport properties with strain, some of them experimentally observed in detail. We also put on the discussion the issue of how accurate it is to describe these systems as j$_{eff}$= 1/2 states. To do this, we provide ab initio calculations based on different schemes together with an oversimplified tight-binding modeling of SrIr$_2$O$_4$.
We can conclude that introducing non-collinearity effects on the Ir magnetic moments is a key ingredient in order to yield a solution that approaches the j$_{eff}$= 1/2 description of the t$_{2g}$ hole,
as well as to capture the weak ferromagentic component.
However, the effects on the actual band structure of introducing a solution that resembles the j$_{eff}$= 1/2 state are very minor, and the evolution with strain is independent of how close to that solution the system is. A solution based on a combination of $xz/yz$ orbitals plus spin-orbit coupling, which is obtained within a collinear description yields
a very similar band structure and the same dependence on strain. However a non-collinear description is essential to determine the magnetic properties of the system and the orbital character of the single t$_{2g}$ unoccupied band.

\acknowledgments

This work was supported by Xunta de Galicia under the Emerxentes Program via the project no. EM2013/037 and the MINECO via project MAT2013-44673-R. V.P. acknowledges support from the MINECO of Spain via the Ramon y Cajal program. J. L. Lado
thanks ITN Spinograph for financial support. We thank W.E. Pickett, P. Blaha and C.D. Batista for insightful comments.



%

\end{document}